\documentclass[apjl]{emulateapj}
\usepackage{apjfonts}

\shorttitle{The widest very low mass binary}
\shortauthors{Artigau et al.}

\begin{document}

\title{Discovery of the widest very low mass binary}

\author{\'Etienne Artigau\altaffilmark{1,2}, David Lafreni\`ere\altaffilmark{2}, Ren\'e Doyon\altaffilmark{2}, Lo\"{\i}c Albert\altaffilmark{3}, Daniel Nadeau\altaffilmark{2}, Jasmin Robert\altaffilmark{2}}
\altaffiltext{1}{Gemini Observatory, Southern Operations Center, Association of Universities for Research in Astronomy, Inc., Casilla 603, La Serena, Chile}
\altaffiltext{2}{D\'epartement de Physique and Observatoire du Mont M\'egantic, Universit\'e de Montr\'eal, C.P. 6128, Succ. Centre-Ville, Montr\'eal, QC, H3C 3J7, Canada}
\altaffiltext{3}{Canada-France-Hawaii Telescope Corporation, 65-1238 Mamalahoa Highway, Kamuela, HI 96743, USA}
\email{eartigau@gemini.edu}

\begin{abstract}
We report the discovery of a very low mass binary system (primary mass $<$0.1~M$_\odot$) with a projected separation of $\sim$5100~AU, more than twice that of the widest previously known system. A spectrum covering the 1-2.5~$\mu$m wavelength interval at $R\sim$1700 is presented for each component. Analysis of the spectra indicates spectral types of M6.5V and M8V, and the photometric distance of the system is $\sim$62~pc. Given that previous studies have established that no more than $1\%$ of very low mass binary systems have orbits larger than 20~AU, the existence of such a wide system has a bearing on very low mass star formation and evolution models.
\end{abstract}


\section{Introduction}
A large fraction of the stars in the Galaxy reside in multiple systems. Generally, higher multiplicity fractions are found as the stellar mass increases, with a binarity fraction reaching $\sim$80$\%$ for B stars \citep{Kouwenhoven2005}. At the other end of the distribution, surveys of very low mass (VLM, mass $<0.1$M$_\odot$) stars and brown dwarfs (BDs), both in the field and in star forming regions, have found a binary fraction of $\sim$20$\%$ \citep{Burgasser2007}; corrections for incompleteness of tight systems could revise this number to a higher value.

 There are important differences between the VLM binary systems and their more massive counterparts, possibly reflecting differences in their formation mechanism. The mass ratio distribution of VLM binaries is strongly peaked at unity \citep{Burgasser2007}, whereas it is more evenly distributed between 0 and 1 for more massive stars. The results of \citet{Duquennoy1991} indicate that around 50\% of solar-type binary systems have separations larger than $\sim$40~AU, and that these systems can have separations as large as $\sim$20~000~AU. On the other hand, small separations are strongly favored among VLM binary systems, less than $1\%$ of which have separations larger than 20~AU \citep{Burgasser2007}.
 
To date, only five VLM systems with separation $> 50$~AU are known, two in the field, Koenigstuhl 1AB (1800~AU) and DENIS J055146.0-443412.2AB (220~AU) \citep{Billeres2005, Caballero2007}, and three in star forming regions: 2MASS J1623361-240221 (212~AU), 2MASS J1101192-773238AB (242~AU), and 2MASS J1622252-240514 (243~AU) \citep{Jayawardhana2006, Close2006, Luhman2004}. A possible binary system having a separation $\sim$1700 AU was recently identified in the $\sigma$~Orionis cluster \citep{Caballero2006}, but this system lacks proper motion (PM) confirmation of physical association. In this Letter we report the identification of 2MASS~J012655.49-502238.8 and 2MASS~J012702.83-502321.1 as part of a VLM binary system (hereafter referred to as 2M0126AB) with a projected separation of $\sim$5100~AU.

\section{Discovery of the system and observations}
The system was found by comparing the positions of the 2MASS point source catalogue \citep{Cutri2003} with those of the $I$-band Digitized Sky Survey images to search for relatively high PM ($\gtrsim$0\farcs1~yr$^{-1}$) BD candidates at high galactic latitudes ($|b|>30^\circ$). Among the sample of candidates found, the two sources mentioned above appeared to share the same PM. This was further confirmed using $J$-band observations obtained on 2006 June~21 with the Observatoire du Mont M\'egantic wide field near-infrared camera CPAPIR (Artigau et al., in preparation), currently installed at the CTIO 1.5-m telescope. The near-infrared colors (Table~\ref{tbl-1}) of both components are indicative of late-M dwarfs \citep{Leggett2002} and their apparent magnitudes place them at a similar distance, suggesting a physical association (see \S\ref{random}).

We obtained near-infrared spectroscopic observations (R$\sim$1700) of 2MASS J012655.49-502238.8 (component A) and 2MASS J012702.83-502321.1 (component B) with the GNIRS spectrograph \citep{Elias2006} at the Gemini South telescope on 2006 August 11 (see Figure~\ref{fig1}) with a 0\farcs3 slit under a 0\farcs7-1\farcs0 seeing. For each component, 15 individual spectra were obtained for a total integration time of 15 minutes, and their median combination has a signal-to-noise ratio of about 30 per resolution element (2 pixels at R$\sim1700$) in the $J$, $H$, and $K$ bands. The A0 star HD8341 was observed for telluric absorption calibration.

\section{Properties of the components}
A summary of all measurements and derived parameters is presented in Table~\ref{tbl-1}. The proper motion of each component was calculated from their positions in the SuperCosmos Sky Survey (SSS) $I$-band catalog \citep{Hambly2001}, the 2MASS PSC, and the CPAPIR observations; the corresponding epochs are 1983.661, 1999.778, and 2006.469. A correction of ($+0\farcs18,-0\farcs10$) was applied to the SSS coordinates to account for a local systematic offset with respect to the 2MASS PSC, as determined from the comparison of $\sim200$ stars located within 15\arcmin\ from 2M0126AB. The astrometry of the CPAPIR observations was calibrated by using the 2MASS PSC as a reference frame, and the relative astrometric uncertainty after calibration was found to be 0\farcs16. The uncertainties on the SSS coordinates (0\farcs15, 0\farcs3) and the 2MASS coordinates (0\farcs07,0\farcs07) were estimated from \citet{Hambly2001} and the 2MASS PSC, respectively. An error-weighted linear regression was performed to obtain the PMs and the corresponding uncertainties. The PM values found for A and B, respectively $131\pm9$ and $135\pm9$~mas/yr in right ascension and $-53\pm15$ and $-47\pm15$~mas/yr in declination, agree within $1\sigma$. Although the available data make it difficult to quantify, one may assume that part of the uncertainty on the individual PMs arises from calibration errors common to both objects, so that the uncertainties on the difference in PM may be smaller than indicated by the above values.

\begin{figure}
\plotone{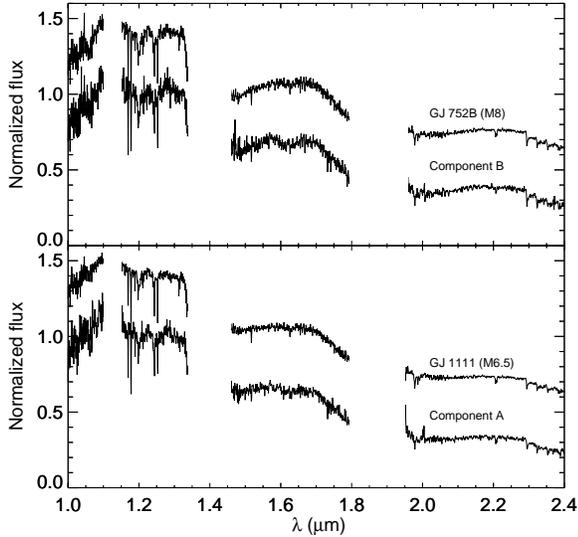}
\caption{\label{fig1} Bottom panel shows the spectrum of component A with the spectrum of an M6.5 dwarf for comparison. The top panel shows the spectrum of B with the spectrum of an M8 dwarf. All spectra are normalized to their mean in the range $1.2$-$1.3$~$\mu$m and the spectra of comparison dwarfs have been offset by $+0.4$. Spectral regions strongly affected by telluric absorption are masked out. The spike-like structure at 2.02~$\mu$m in the spectrum of both A and B is most likely due to an imperfect correction of a narrow telluric absorption band. The comparison spectra are from Rayner et al. (in preparation) and were obtained with SpeX at the IRTF.}
\end{figure}

The spectral types were derived primarily from cross-correlation of both spectra with those of known M5-M9 dwarfs obtained with SpeX at the IRTF\footnote[4]{http://irtfweb.ifa.hawaii.edu/~spex/spexlibrary/IRTFlibrary.html.} (Rayner et al., in preparation). The wavelength intervals used for the cross-correlation were: $1.0-1.12$, $1.18-1.35$, $1.48-1.8$, and $1.95-2.4$~$\mu$m. Spectral types of M6.5-M7 and M8 were found for components A and B respectively  (see Figure~\ref{fig1}). The spectral indices defined in \citet{McLean2003} yield spectral types of M$6.5$-M$9$ and M$6.5$-L$0.5$ for A and B respectively, consistent with the previous determination. The relatively short wavelength intervals (0.004~$\mu$m) used for calculating these indices combined with the modest signal-to-noise ratio of our spectra explain the larger uncertainty on the spectral types using this scheme. The equivalent widths (EWs) of the K\textsc{I} lines at 1.169, 1.177, 1.244, and 1.253~$\mu$m for components A and B (see Table~\ref{tbl-1}) are consistent with those of M6-M6.5 and M7-M9 dwarfs respectively, based on Figure 14 and 15 of \citet{McLean2003}. Considering all of these measurements, we adopt a spectral type of M6.5V for A and M8V for B with an uncertainty of $\pm0.5$.

\begin{table}
\begin{center}
\caption{Parameters of 2MASS J0126AB\label{tbl-1}}
\begin{tabular}{lcc}
\hline\hline
Parameter & A & B\\
\hline
2MASS designation&J012655.49-502238.8&J012702.83-502321.1\\
Angular separation&\multicolumn{2}{c}{81\farcs86$\pm$0\farcs10}\\
$\mu_\alpha \cos\delta^{\rm a}$&$131\pm9$~mas/yr&$135\pm9$~mas/yr\\
$\mu_\delta^{\rm a}$&$-53\pm$15 mas/yr&$-47\pm$15 mas/yr\\
$V^{\rm b}$&21.8&22.2\\
$I^{\rm c}$&17.2$\pm$0.3&17.6$\pm$0.3\\
$J^{\rm d}$&14.61$\pm$0.04&14.81$\pm$0.05\\
$H^{\rm d}$&14.05$\pm$0.05&14.16$\pm$0.04\\
$K_s^{\rm d}$&13.68$\pm$0.05&13.62$\pm$0.05\\
FeH 1.20~$\mu$m$^{\rm e}$&$6.0\pm1.8$~\AA& $14.5\pm1.9$~\AA\\
K\textsc{I}~1.25$\mu$m$^{\rm e}$&$9.8\pm1.1$~\AA&$10.3\pm1.0$~\AA\\
Spectral type&M6.5V $\pm$ 0.5&M8V $\pm$ 0.5\\
Photometric distance$^{\rm f}$&63$\pm$5 pc&61$\pm$6 pc\\
Physical separation&\multicolumn{2}{c}{5100$\pm$400~AU}\\
$T_{eff}^{\rm g}$&$2670\pm180$~K&$2490\pm180$~K\\
Mass ($>1$~Gyr)$^{\rm h}$&$0.095\pm0.005$~M$_{\odot}$&$0.092\pm0.005$~M$_{\odot}$\\
Mass ($\sim$30~Myr)$^{\rm h}$&$0.020\pm0.003$~M$_{\odot}$&$0.019\pm0.003$~M$_{\odot}$\\
\hline\hline
\end{tabular}
\tablenotetext{a}{Calculated using the SuperCosmos Sky Survey $I$-band positions (epoch 1983.661), the 2MASS point source catalogue position (epoch 1999.778), and our measurements made with CPAPIR (epoch 2006.469).}
\tablenotetext{b}{Assuming $V-I = 4.6$ for both objects (see \citealp{Dahn2002}, fig~$3$)}
\tablenotetext{c}{From the SuperCosmos Sky Survey catalogue.}
\tablenotetext{d}{From the 2MASS point source catalogue.}
\tablenotetext{e}{Spectral indices defined in \citet{Gorlova2003}}
\tablenotetext{f}{Determined by comparing the $J$, $H$, and $K_s$ magnitudes of A and B with the absolute magnitudes of eight M6.5V-M7V and four M8V dwarfs with known trigonometric parallaxes; the quoted uncertainties correspond to the dispersions of the distances obtained.}
\tablenotetext{g}{Spectral-type $-$ $T_{eff}$ relation of \citet{Golimowski2004}.}
\tablenotetext{h}{Determined by minimizing the sum of the square difference between the absolute $J$, $H$, and $K_s$ magnitudes of 2M0126AB and evolution models \citep{Chabrier2000}. The quoted uncertainties are derived from uncertainties in the photometric distance.}
\end{center}
\end{table}

Photometric distances of $63\pm5$~pc and $61\pm6$~pc are calculated for A and B respectively from comparison of their $JHK_s$ magnitudes with the absolute magnitudes of other dwarfs of the same spectral types with known trigonometric parallaxes. Using the spectral type$-$effective temperature ($T_{eff}$) relation of \citet{Golimowski2004}, we infer $T_{eff} =  2670\pm180$~K for A and $T_{eff} =  2490\pm180$~K for B. According to evolution models \citep{Chabrier2000}, such temperatures imply masses below 0.1~M$_\odot$, irrespective of the system age.

\section{Probability of random alignment}
\label{random}
Assuming a late-M dwarf spatial density of $2.2\times10^{-3}$~pc$^{-3}$ per $I$-band magnitude interval \citep{Phan-Bao2003}, there should be $\sim$2700 M6-M8 dwarfs ($12.5<$M$_I<14.5$) between 55 and 68~pc from the Sun. Considering that an 82$^{\prime\prime}$ radius disk occupies $3.95\times10^{-8}$ of the sky, and assuming an isotropic distribution of late-Ms out to $68$~pc, the probability of finding at least one such pair of objects over the whole sky from chance alignment is $0.15$, regardless of their apparent proper motion. A Monte Carlo calculation was performed to determine the probability that two M dwarfs located in the direction of 2M0126AB have PMs larger than 0\farcs1 yr$^{-1}$ that agree with each other within our uncertainties at the 2~$\sigma$ level. Using the observed dispersions in ($U$, $V$, $W$) of late-M dwarfs in the Solar neighborhood \citep{Bochanski2005}, including the disk and halo components, and converting these velocities into PMs at the position of 2M0126AB, the probability is $1.4\times10^{-2}$. Thus, the combined probability of finding a pair of late-M dwarfs within 82$^{\prime\prime}$ of each other and sharing the same proper motion within our uncertainties is 2$\times10^{-3}$. Therefore, 2M0126AB is very likely a bound system rather than a random alignment of two unrelated objects.

\section{Discussion}\label{discussion}
As shown in \citet{Gorlova2003}, the EWs of FeH and K\textsc{I} provide a rough constraint on surface gravity. Indeed, the strength of these features in late-type M dwarfs is systematically higher (by a factor 1.5-2.0) for field objects compared to young ($\lesssim10$~Myr) cluster members. Based on Figure~8 of \citet{Gorlova2003}, the measured FeH 1.20$\mu$m and K\textsc{I} 1.25$\mu$m EWs of components A and B are in better agreement with those of relatively old field objects. The mass of each component can be determined from comparison of near-infrared absolute magnitudes with evolution models \citep{Chabrier2000} assuming an age for the system. If we suppose an age greater than 1 Gyr, as suggested by the gravity sensitive spectral indices, we derive masses of $0.095\pm0.005$ and $0.092\pm0.005$~M$_\odot$ for the two components.

\begin{figure}
\plotone{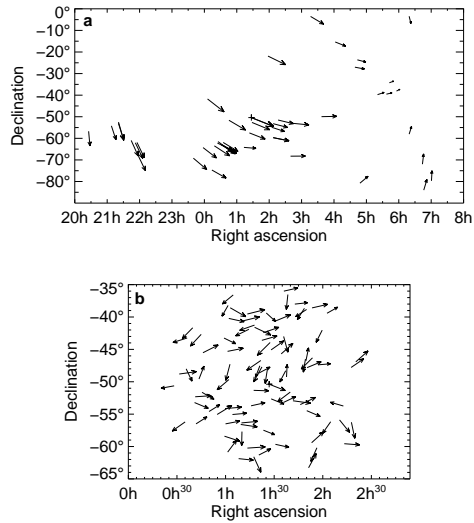}
\caption{\label{fig2} (a) The PM of all TH members listed in \citet{Zuckerman2004} and of 2M0126AB are indicated by an arrow originating from their right ascension and declination. The position of 2M0126AB is indicated by a cross. 2M0126AB lies in the core of TH and its tangential motion is very similar to that of nearby members. (b) Position and PM of 2M0126AB and of all stars in the Hipparcos catalogue located within 15 degrees of 2M0126AB and with $100<$ PM (mas/yr) $<150$. The position of 2M0126AB is indicated by a cross. There is no overall tendency for stars in this part of the sky to move in a direction similar to that of 2M0126AB, or of members of TH.}
\end{figure}

Although the FeH and K\textsc{I} EWs are consistent with relatively old field objects, the age of 2M0126AB remains poorly constrained given the small difference in EW between field and very young dwarfs and the significant dispersion of those values (see Figure~8 of \citealp{Gorlova2003}). In fact, there is some evidence that 2M0126AB might be part of a young association. On the sky, it lies in the core of the young ($\sim$30~Myr) Tucana/Horologium (TH) association (\citealp{Zuckerman2004} and references therein) located at a distance of 37-66 pc, consistent with that estimated for 2M0126AB. Furthermore, both the amplitude and direction of its PM are similar to those of nearby members (see Figure~\ref{fig2}). We note, however, that the amplitude of the PM of 2M0126AB (142~mas/yr) is slightly larger than the largest PM of nearby members (124~mas/yr). The PM dispersion of members of the association is $\sim$11~mas/yr, and is representative of relatively massive members ($\gtrsim$0.5 M$_\odot$). The dispersion is expected to be significantly higher for lower mass members (see Figure~4 of \citealp{Pinfield1998}).  An age of 30 Myr for 2M0126AB would imply that the two components are BDs with masses of $0.020\pm0.003$ and $0.019\pm0.003$~M$_\odot$. Future determinations of the trigonometric parallax, the radial velocity, the lithium abundance, the surface gravity, and the H$\alpha$ or \hbox{X-ray} emission of 2M0126AB will help in settling this issue, and in refining the physical characterization of the system.

In the scenario where 2M0126AB is a member of the TH association, the random alignment probability determined earlier needs to be re-evaluated; the major argument in favor of a physical association now being the relatively small angular separation of the system given the low stellar density of the association. The association has a diameter of approximately 30 pc, hence a volume of $1.4\times10^4$~pc$^3$. The uncertainties on the distance of both components and their angular separation place them within a volume $5.7\times10^{5}$ times smaller. Assuming a Kroupa three-segment IMF \citep{Kroupa2001}, and knowing that there are 42 members of the TH association between B8 and K5 (roughly 4~M$_\odot$ to 0.67~M$_\odot$), we estimate that the association comprises 32 objects in the mass interval between $0.015~$M$_\odot$ and $0.025~$M$_\odot$. Given these numbers, the probability that the two components are not physically bound is only \hbox{$8.7\times10^{-4}$}. Even assuming a factor of 5 underestimate of the number of low-mass objects in the association, this probability would still be only 2~$\%$.

\begin{figure}
\plotone{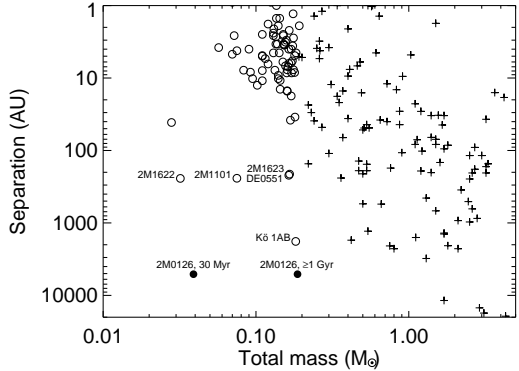}
\caption{\label{fig3} Orbital separation vs total mass of binary systems. VLM systems are shown with empty circles and stellar systems are shown with crosses. The left and right filled circles show 2M0126AB if it is aged 30 Myr (i.e. a member of TH) or $> 1$ Gyr respectively. The labels refer to the large separation VLM systems mentioned in the text.}
\end{figure}

At a distance of 62~pc, the measured angular separation of the system corresponds to a projected physical separation of $\sim$5100~AU, more than twice that of the Koenigstuhl~1AB system \citep{Caballero2007} and $\sim$20~times greater than those of other known wide VLM binaries (see Figure~\ref{fig3}). Because of projection effects, the real separation is expected to be, on average, 1.4 times larger \citep{Couteau1960} which, in the present case, would be $\sim$7000 AU. Prior to the discovery of 2M0126AB, projected separations of $\sim$5000~AU had been observed only for systems of total mass $>1$~M$_\odot$. The low total mass and large separation of 2M0126AB raise the question of its dynamical stability. An order of magnitude estimate of the expected survival time of 2M0126AB can be obtained from the work of \citet{Weinberg1987}. Stellar and giant molecular cloud (GMC) encounters are the dominant source of disruption for low-mass binaries, and these contributions are dependent on the mass/separation ratio only, implying that the results derived for more massive binaries can be scaled to the 2M0126AB system. Given this scaling relation and Figure~2 of \citet{Weinberg1987}, the stellar and GMC encounter disruption timescale of 2M0126AB is $\sim$200~Myr if it is a member of the 30 Myr old TH association or $\sim$1~Gyr if it is a field object. The first timescale above is well beyond the age of the TH association, while the second may indicate that 2M0126AB is likely younger than 2-3 Gyr. Further, the second timescale above is long enough for a significant number of similar binaries produced throughout the history of the Galaxy to have survived to this day.

The formation process of VLM stars and BDs is the subject of an active debate, partly because their mass is less than $\sim$0.7~M$_\odot$, the typical Jeans mass in molecular clouds \citep{Meyer2000}. However, it has been shown recently \citep{Boyd2005, Padoan2004, Padoan2005} that turbulence can lead to fragmentation of unstable cores with masses as small as $\sim$0.003~M$_\odot$.  A possibility is that VLM stars and BDs are such low-mass embryos dynamically ejected out of their feeding zone during the early stages of formation \citep{Reipurth2001}. This is a popular scenario because it agrees with the observation that most VLM binaries have close separations, weakly bound wider binaries being easily disrupted by the ejection. Nevertheless, detailed simulations \citep{Bate2005} of this mechanism have succeeded in producing some VLM binaries with separations $\sim$100~AU, similar to the four previously known VLM binaries mentioned above albeit with lower mass ratios. These wide binaries result when low mass cores become bound following ejection from the cluster at the same time and in the same direction. It remains to be seen whether binaries like 2M0126AB, with orbital separations $\sim$20~times larger, can be produced by this mechanism.

Photo-evaporation of pre-stellar cores by the strong ionizing radiation of nearby massive stars \citep{Whitworth2004} is another possible VLM star and BD formation process. However, the presence of massive stars required for this to happen would most likely disrupt a VLM binary such as the one described here. Formation of the secondary by direct gravitational collapse within the accretion disk of the primary \citep{Boss2000, Boss2006} is yet another possibility, but it is unlikely to form nearly equal mass binaries and would require a disk of unreasonably large spatial extent.

The formation of BDs following the encounter of proto-stellar disks has been shown possible in simulations \citep{Shen2006}. In this case, the disks are normally stable against gravitational collapse; it is only the violent collision between two disks that triggers instabilities leading to the formation of BDs. Although current simulations are not sufficient to address directly the question of wide binary BD formation, the results indicate that about one third of the BDs produced are ejected from the system, and wide binaries may result as in the ejection scenario discussed above. So far, the simulations have only produced BD mass objects and not VLM stars through this mechanism.

\acknowledgments
The authors are grateful to the anonymous referee for helpful comments that improved the quality of the paper. This work was supported in part through grants from the Natural Sciences and Engineering Research Council, Canada, and from Fonds Qu\'eb\'ecois de Recherche sur la Nature et les Technologies. This publication has made use of the VLM Binaries Archive maintained by Nick Siegler at http://paperclip.as.arizona.edu/$\sim$nsiegler/VLM\_binaries/. Based on observations obtained at the Gemini Observatory, which is operated by the Association of Universities for Research in Astronomy, Inc., under a cooperative agreement with the NSF on behalf of the Gemini partnership: the National Science Foundation (United States), the Particle Physics and Astronomy Research Council (United Kingdom), the National Research Council (Canada), CONICYT (Chile), the Australian Research Council (Australia), CNPq (Brazil) and CONICET (Argentina). The Digitized Sky Survey was produced at the Space Telescope Science Institute under U.S. Government grant NAG W-2166. The images of these surveys are based on photographic data obtained using the Oschin Schmidt Telescope on Palomar Mountain and the UK Schmidt Telescope.

\end{document}